\documentclass[a4paper,11pt]{article}
\usepackage[utf8]{inputenc}
\usepackage{amsmath,amssymb}
\usepackage{slashed}
\usepackage{graphicx}
\usepackage{cite}
\usepackage{hyperref}
\usepackage{chngcntr}
\usepackage{subcaption}
\usepackage{braket,slashed}
\usepackage{lipsum}

\newcommand\blfootnote[1]{%
	\begingroup
	\renewcommand\thefootnote{}\footnote{#1}%
	\addtocounter{footnote}{-1}%
	\endgroup
}

\allowdisplaybreaks

\begin{document}
	
	\begin{titlepage}
		
		\vspace*{-2cm}

		\begin{center}
			\vspace*{15mm}
			
			\vspace{1cm}
			{\LARGE \bf
				Structure-Dependent Electromagnetic \\ \vspace{0.2cm} Finite-Size Effects
			} 
			\vspace{0.8cm}
			
			\renewcommand{\thefootnote}{\fnsymbol{footnote}}
			{Matteo Di Carlo$^a$, Maxwell T.~Hansen$^a$, Nils Hermansson-Truedsson$^{b,*}$\blfootnote{* Speaker}, Antonin Portelli$^a$}
			\renewcommand{\thefootnote}{\arabic{footnote}}
			\setcounter{footnote}{0}
			
			\vspace*{.5cm}
			\centerline{$^a${\it Higgs Centre for Theoretical Physics, 
					School of Physics and Astronomy,}} 
			\centerline{\it  University of Edinburgh, Peter Guthrie Tait Road,
				Edinburgh,
				EH9 3FD, United Kingdom}
								\vspace{0.2cm}
			\centerline{${}^b$ \it Albert Einstein Center for Fundamental Physics, Institute for Theoretical Physics, }
			\centerline{{\it Universit\"{a}t Bern, Sidlerstrasse 5, CH–3012 Bern, Switzerland}}
			\vspace*{.2cm}
			
		\end{center}
		
		\vspace*{10mm}
		\begin{abstract}\noindent\normalsize
		We present a model-independent and relativistic approach to analytically derive electromagnetic finite-size effects beyond the point-like approximation. The key element is the use of electromagnetic Ward identities to constrain vertex functions, and structure-dependence appears via physical form-factors and their derivatives. We apply our general method to study the leading finite-size structure-dependence in the pseudoscalar mass (at order $1/L^3$) as well as in the leptonic decay amplitudes of pions and kaons (at order $1/L^2$). Knowledge of the latter is essential for Standard Model precision tests in the flavour physics sector from lattice simulations.  
		\end{abstract}
	
	\vspace{2cm}
Proceedings for talks presented at 	
\begin{itemize}
	\item[] \textit{The 38th International Symposium on Lattice Field Theory, LATTICE2021,
		July 26-30 2021  }
	\item[] \textit{A Virtual Tribute to Quark Confinement and the Hadron Spectrum 2021, August 2-6 2021}
	\item[] \textit{Particles and Nuclei International Conference 2021, PANIC2021, September 5-10 2021}
\end{itemize}
		
	\end{titlepage}
	\newpage

\section{Introduction}

Lattice quantum chromodynamics (QCD) allows for systematically improvable Standard Model (SM) precision tests from numerical simulations performed in a finite-volume (FV), discretised Euclidean spacetime. In order to reach (sub-)percent precision in lattice predictions, also strong and electromagnetic isospin breaking corrections have to be included. The latter are encoded via quantum electrodynamics (QED), but the inclusion of QED in a FV spacetime is complicated because of Gauss' law~\cite{Hayakawa:2008an}. This problem is related to zero-momentum modes of photons and the absence of a QED mass-gap. Several prescriptions of how to include QED in a finite volume have been formulated and the one used here is QED$_{\textrm{L}}$ where the spatial zero-modes are removed on each time-slice. The long-range nature of QED in addition enhances the FV effects (FVEs), which typically leads to power-law FVEs that are larger than the exponentially suppressed ones for single-particle matrix elements in QCD alone. 

The FVEs for a QCD+QED process depend on properties of the involved particles, including masses and charges, but also structure-dependent quantities such as electromagnetic form-factors and their derivatives. In order to analytically capture the finite-volume scaling fully, one cannot neglect hadron structure, and in the following we develop a relativistic and model-independent method to go beyond the point-like approximation at order $e^2$ in QED$_{\textrm{L}}$. 

We consider a space-time with periodic spatial extents $L$ but with infinite time-extent. To exemplify the method, we first consider the pseudoscalar mass in Sec.~\ref{sec:se}, and then proceed to leptonic decays in Sec.~\ref{sec:kl2}. The discussion is based on the results in Ref.~\cite{DiCarlo:2021apt}, and the reader is referred there for further technical details. 

\section{Pseudoscalar Mass}\label{sec:se}
To study the finite-size scaling in the mass $m_{P}(L)$ of a charged hadronic spin-0 particle $P$, we first define the full QCD+QED infinite-volume two-point Euclidean correlation function
\begin{align}
	C^\infty_2(p) & =\int d^4 x\,\bra{0}\mathrm{T}[\phi(x)\phi^\dagger(0)]\ket{0}
	e^{-ipx}\,.\label{eq:c2def}  
\end{align}
Here $\phi$ is an interpolating operator coupling to $P$, and $p=(p_{0},\mathbf{p})$ is the momentum. We denote the finite-volume counterpart of this correlator $C^L_2(p)$, but for the moment only consider $C^\infty_2(p)$. This can be diagrammatically represented as
\begin{equation}\label{eq:qcdqed2pointdiag}
	C^\infty_2(p)=\raisebox{-1.3ex}{\includegraphics{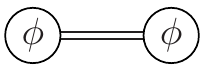}}  = Z_{P} \cdot D(p) \cdot Z_{P}\,,  
	\qquad  
	D(p) = \frac{Z(p^2)}{p^2 + m_P^2} \, ,
	\qquad 
	Z_P = \bra{0}\phi(0)\ket{P,\mathbf{p}}\,,
\end{equation}
where the double-line represents the QCD+QED propagator $D(p)$, the $\phi$-blob is the overlap between $\phi$ and $P$ and  $Z(p^2)=1+\mathcal{O}(p^2+m_{P}^2)$ is the residue of the propagator. Expanding $C^\infty_2(p)$ in~(\ref{eq:qcdqed2pointdiag}) around $e=0$ yields
\begin{equation}
	\raisebox{-1.3ex}{\includegraphics{axo_full2pt.pdf}}=
	\raisebox{-1.3ex}{\includegraphics{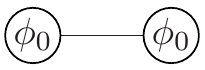}}+
	\raisebox{-2.1ex}{\includegraphics{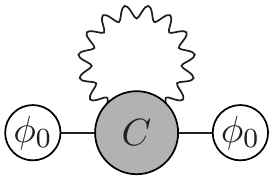}}+\mathcal{O}(e^4)\,,
	\label{eq:2ptexp}
\end{equation}
where quantities with subscript $0$ are evaluated in QCD alone. The grey blob is the Compton scattering kernel defined via
\begin{align}\label{eq:comptonkernelcorrfcn}
	\raisebox{-2.1ex}{\includegraphics{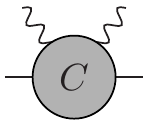}} = C_{\mu\nu}(p,k, q) = 
	\int d^4 x \, d^4 y \,d^4 z \, e^{i p z + i k x + i q y}  \frac{\bra{0}\mathrm{T}[\phi(0) J_\mu(x)  J_\nu(y) \phi^\dagger(z)]\ket{0}}{Z^{2}_{P,0} \, D_0(p)  D_0(p + k + q) }
	\,.
\end{align}
Here $k$ and $q$ are incoming photon momenta and $J_\mu (x)$ is the electromagnetic current. Note that the unphysical dependence on the arbitrary interpolating operator $\phi$ must cancel for any physical quantity, and when the external legs in $C_{\mu\nu}(p,k, q)$ go on-shell the kernel is nothing but the physical forward Compton scattering amplitude. Using~(\ref{eq:2ptexp}) the electromagnetic mass-shift of the meson is readily obtained in terms of an integral over the photon loop-momentum $k$. One may follow an equivalent procedure for the finite-volume correlation function $C^L_2(p)$, where the integral over the spatial momentum $\mathbf{k}$ is replaced by a sum. The leading electromagnetic FVEs in the mass, $\Delta m_{P}^2(L)$, are thus given by the sum-integral difference
\begin{align}\label{eq:fvemassgeneral}
	\Delta m^2_P(L) & \,  
	= - \frac{e^2}{2} \lim_{p_0^2 \to - m_{P}^2}     \left( \left. \frac{1}{L^3}\sum_{\mathbf{k} }\right. '-\int\frac{d ^3\mathbf{k}}{(2\pi)^3} \right)
	\int\frac{d k_0}{ 2\pi }
	\left. \frac{C_{\mu\mu}(p,k,-k)}{k^2}  \right| _{\mathbf{p} =  0}\,,
\end{align}
where the rest-frame $\mathbf{p}=0$ was chosen for convenience and the primed sum indicates the omission of the photon zero-mode $\mathbf{k}=0$ in QED$_{\textrm{L}}$. The analytical dependence on $1/L$ including structure-dependence can now be obtained from this formula through a soft-photon expansion of the integrand, i.e.~an expansion order by order in $|\mathbf{k}|$ which is directly related to the expansion in $1/L$ via $|\mathbf{k}|= 2\pi |\mathbf{n}|/L$ where $\mathbf{n}$ is a vector of integers.  
The first step is to decompose $C_{\mu\nu}(p,k,q)$ into two irreducible electromagnetic vertex functions $\Gamma_{1}$ and $\Gamma _2$ according to
\begin{equation}\label{eq:skelexpcompton}
	\raisebox{-2.1ex}{\includegraphics{axo_sigker.pdf}}=
	\raisebox{-2.1ex}{\includegraphics{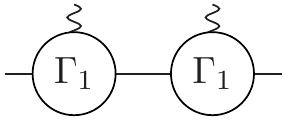}}+\raisebox{-2.1ex}{\includegraphics{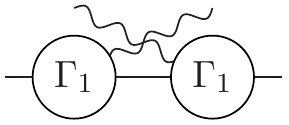}}+
	\raisebox{-2.1ex}{\includegraphics{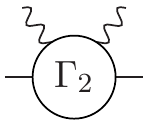}}\,.
\end{equation}
The vertex functions depend in general on the structure of the particle, as can be seen from e.g.~the form-factor decomposition
\begin{equation}
	\label{eq:Gammamu}
	\Gamma _{1} =  \Gamma_\mu(p,k) = (2p+k)_\mu \,  F(k^2,(p+k)^2,p^2) + k_\mu \,  G(k^2,(p+k)^2,p^2)\,,
\end{equation}
where $F(k^2,(p+k)^2,p^2)$ and $G(k^2,(p+k)^2,p^2)$ are structure-dependent electromagnetic form-factors depending on three virtualities. This means that $F$ and $G$ contain off-shell effects, but we stress that these non-physical quantities always cancel in the FVEs. The cancellation occurs since the vertex functions $\Gamma_{1,2}$ are related to each other and the propagator $D_{0}(p)$ via Ward identities. An example of an off-shell relation is $F(0,p^2,-m_{P,0}^2)=Z_{0}(p^2)^{-1}$. The derivatives of $Z_{0}(p^2)$ are already known in the literature as $\delta D^{(n)}(0)$~\cite{Borsanyi:2014jba} and $z_{n}$~\cite{Lubicz:2016xro}, but these could in principle be set to zero as they always cancel in the final results. The Ward identities further yield $G$ as a function of $F$. The form-factor $F$ also contains physical information, and for our purposes it suffices to know that $F^{(1,0,0)}(0,-m_{P,0}^2,-m_{P,0}^2) = F'(0) = -\braket{r_P^2}/6$, where $\braket{r_P^2}$ is the physical electromagnetic charge radius of $P$ which is well-known experimentally~\cite{10.1093/ptep/ptaa104}. 

Using our definitions of the vertex functions in $C_{\mu\nu}(p,k,q)$ in~(\ref{eq:fvemassgeneral}) we obtain the FVEs
\begin{align}\label{eq:fveffmass}
	\Delta m_P^2(L) = e^2 m_P^2\left\{\frac{c_{2}}{4\pi^2m_P L} +\frac{c_{1}}{2\pi(m_P L)^2} + \frac{\braket{r_P^2}}{3m_PL^3} + \frac{\mathcal C}{(m_P L)^3} +\mathcal{O}\left[\frac{1}{(m_PL)^4}\right]\right\}
	\,,
\end{align}
where the $c_{j}$ are finite-volume coefficients specific to QED$_{\textrm{L}}$ arising from the sum-integral difference in~(\ref{eq:fvemassgeneral}). These are discussed in detail in Ref.~\cite{DiCarlo:2021apt}. Here we see the charge radius $\braket{r_P^2}$ appearing at order $1/L^3$ and its coefficient agrees with that derived within non-relativistic scalar QED~\cite{Davoudi:2014qua}. However, there is an additional structure-dependent term $\mathcal{C}$ related to the branch-cut of the forward, on-shell Compton amplitude. This contribution can be found, in other forms, also in Refs.~\cite{Borsanyi:2014jba,Tantalo:2016vxk}, and only arises because of the QED$_{\textrm{L}}$ prescription with the subtracted zero-mode. Its value is currently unknown but one can show $\mathcal{C}>0$~\cite{DiCarlo:2021apt}, meaning that it cannot cancel the charge radius contribution. Note that all unphysical off-shell contributions from the form-factors $F$ and $G$ have vanished.

\section{Leptonic Decays}\label{sec:kl2}
Leptonic decay rates of light mesons are of the form $P^{-}\rightarrow
\ell ^{-}\bar{\nu} _{\ell}$, where $P$ is a pion or kaon, $\ell $ a lepton and $\nu _{\ell}$ the corresponding neutrino. These are important for the extraction of the Cabibbo-Kobayashi-Maskawa matrix elements $|V_{us}|$ and $|V_{ud}|$~\cite{Carrasco:2015xwa,DiCarlo:2019thl}. The leading virtual electromagnetic correction to this process yields an infrared (IR) divergent decay rate $\Gamma _{0}$. One must therefore add the real radiative decay rate $\Gamma _{1}(\Delta E)$ for $P^{-}\rightarrow
\ell ^{-}\bar{\nu} _{\ell}\gamma $, where the photon has energy below $\Delta E$, to cancel the IR-divergence in $\Gamma _{0}$. The IR-finite inclusive decay rate is thus $\Gamma \left(  P^{-}\rightarrow \ell ^{-}\nu _{\ell}[\gamma ]\right)$, and following the lattice procedure first laid out in Ref.~\cite{Carrasco:2015xwa} we may write
\begin{align}
	& 
	\Gamma _{0}+\Gamma _{1}(\Delta E_{\gamma})   =  \lim _{L\to \infty} [ \Gamma _{0} (L) - \Gamma _{0}^{\mathrm{uni}} (L) ] +\lim _{L\to \infty} [\Gamma _{0}^{\mathrm{uni}} (L) +\Gamma _{1}(L, \Delta E_{\gamma}) ]\label{eq:gammadec}
	\, .
\end{align}
Here, Ref.~\cite{Carrasco:2015xwa} chose to add and subtract the universal finite-volume decay rate $\Gamma ^{\textrm{uni}}(L)$, calculated in point-like scalar QED in Ref.~\cite{Lubicz:2016xro}, to cancel separately the IR-divergences in $\Gamma_{0}$ and $\Gamma _{1}$. In the following we are interested in only the first term in brackets. The subtracted term $\Gamma _{0}^{\mathrm{uni}} (L)$ cancels the FVEs in $\Gamma _{0}(L)$ through order $1/L$, and hence $\Gamma _{0} (L) - \Gamma _{0}^{\mathrm{uni}} (L)\sim \mathcal{O}(1/L^2)$. Structure-dependence enters at order $1/L^2$. With the goal of systematically improving the finite-volume scaling order by order including structure-dependence, we replace the universal contribution by 
\begin{align}\label{eq:decratefve}
	\Gamma _{0}^{\mathrm{uni}} (L) \longrightarrow \Gamma _{0}^{(n)} (L) = \Gamma _{0}^{\mathrm{uni}} (L) + \sum _{j=2}^{n}\Delta \Gamma _{0}^{(j)}(L) \, ,
\end{align}
where $n\geq 2$ and $\Delta \Gamma _{0}^{(j)}(L)$ contains the FVEs at order $1/L^{j}$. This means that the finite-volume residual instead scales as $\Gamma _{0} (L) - \Gamma _{0}^{(n)} (L)\sim \mathcal{O}(1/L^{n+1})$. We may parametrise $\Gamma _{0}^{(n)} (L)$ in terms of a finite-volume function $Y^{(n)}(L) $ according to 
\begin{align}\label{eq:y2def}
	\Gamma _{0}^{(n)}(L)  = \Gamma _{0}^{\textrm{tree}} \left[ 1+2 \frac{\alpha }{4\pi } \, Y^{(n)}(L) \right]+\mathcal{O} \left( \frac{1}{L^{n+1}} \right)
	\, ,
\end{align}
where $\Gamma _{0}^{\textrm{tree}}$ is the tree-level decay rate. 

Since we are interested in the leading structure-dependent contribution we consider $Y^{(2)}(L)$. In order to derive it, we define the QCD+QED correlation function
\begin{align}
	\label{eq:CWfull}
	C_{W}^{rs}(p,p_{\ell}) = \int d^{4}z\, e^{ip z}\,\bra{\ell^-,\mathbf{p}_\ell,r;\nu_{\ell},\mathbf{p}_{\nu _{\ell}},s}\mathrm{T}[\mathcal{O}_{W}(0)\phi^{\dagger}(z)]\ket{0}\, ,
\end{align}
where $p_{\ell} = (p^{0}_{\ell},\mathbf{p}_{\ell})$ is the momentum of the on-shell lepton of mass $m_{\ell}$, $p_{\nu_{\ell}} = (p^{0}_{\nu_{\ell}},\mathbf{p}_{\nu_{\ell}})$ is the momentum of the massless neutrino and $\mathcal{O}_{W}(0)$ is the four-fermion operator of the decay in question. We may diagrammatically represent this in a similar way as for the mass according to  
\begin{equation}\label{eq:cwfull}
	C_W^{rs}(p,p_\ell)=\raisebox{-3.7ex}{\includegraphics{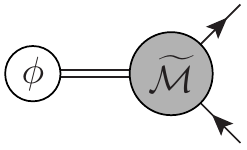}} = \raisebox{-3.7ex}{\includegraphics{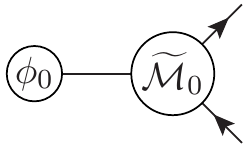}}+\raisebox{-3.7ex}{\includegraphics{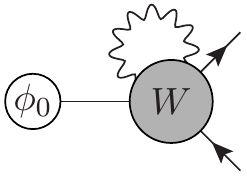}}\,.
\end{equation}
The grey blob containing $W$ is of order $e^2$ and can be separated, just like the Compton amplitude, into several irreducible vertex functions. The exact definitions of these vertex functions are quite involved and can be found in Ref.~\cite{DiCarlo:2021apt}, but several comments can be made. First of all, the vertex functions are related to various structure-dependent form-factors containing both on-shell and off-shell information. Again, the off-shellness must cancel. The vertex functions also contain physical structure-dependent information (similar to how $\Gamma _{1}$ depends on the charge radius) and for $Y^{(2)}(L)$ this is the axial-vector form-factor $F_{A}(-m_{P}^2)=F_{A}^{P}$ from the real radiative decay $P^{-}\rightarrow \ell ^{-}\bar{\nu}_{\ell}\gamma$. 

By performing the amputation on the external meson leg in~(\ref{eq:CWfull}) to obtain the matrix element needed for the decay rate in~(\ref{eq:y2def}), one finds the finite-volume function $Y^{(2)}(L)$ to be
\begin{align} 
	\label{eq:y2finalform}
	Y^{(2)}(L) \, &= \ \frac34 + 4\, \log\left(\frac{m_\ell}{m_W}\right) + 2\,\log\left(\frac{m_W L}{4\pi}\right) \ +\frac{c_3-2\,( c_3(\mathbf{v}_\ell)-B_1(\mathbf{v}_\ell))}{2\pi}\, - \\
	& 
	- 2\,A_1(\mathbf{v}_\ell)\left[\log\left(\frac{m_P L}{2\pi}\right)+\log\left(\frac{m_\ell L}{2\pi}\right)-1\right]  - \frac{1}{m_P L} \left[ \frac{(1+r_\ell^2)^2\,c_2 - 4\, r_\ell^2 \, c_2(\mathbf{v}_\ell)}{ 1-r_\ell^4} \right] + \nonumber\\
	& + \, \frac{1}{(m_P L)^2} \left[ - \frac{F_A^{P}}{f_P} \, \frac{4\pi \, m_P \,[(1+r_\ell^2)^2 \, c_1 - 4\, r_\ell^2\, c_1(\mathbf{v}_\ell)]}{1-r_\ell^4} + \frac{8\pi \, [(1+r_\ell^2) \, c_1 - 2\,  c_1(\mathbf{v}_\ell)]}{ (1-r_\ell^4)} \right]\,. \nonumber
\end{align}
Here, $r_{\ell}=m_{\ell}/m_{P}$, $\mathbf{v}_{\ell}=\mathbf{p}_{\ell}/E_{\ell}$ the lepton velocity in terms of the energy $E_{\ell}$, and $m_{W}$ the $W$-boson mass. Also, $c_{k}$, $A_{1}(\mathbf{v}_{\ell})$, $B_{1}(\mathbf{v}_{\ell})$ and $c_{j}(\mathbf{v}_{\ell})$ are finite-volume coefficients defined in Ref.~\cite{DiCarlo:2021apt}. Note that no unphysical quantities appear. At order $1/L^2$, there is one structure-dependent contribution proportional to $F_{A}^{P}$ and the other term is purely point-like. This result is in perfect agreement with Ref.~\cite{Lubicz:2016xro} for the universal terms up to $\mathcal{O}(1/L)$, which we derived in a completely different approach. The numerical impact of the $1/L^2$-corrections is studied in Ref.~\cite{DiCarlo:2021apt}.

\section{Conclusions}
We have presented a relativistic and model-independent method to derive electromagnetic FVEs beyond the point-like approximation. 
We are currently working to obtain the leading FVEs for semi-leptonic kaon decays, relevant for future precision tests in the SM flavour physics sector.

\section*{Acknowledgements}

M.D.C, M.T.H., and A.P. are supported in part by UK STFC grant ST/P000630/1. Additionally M.T.H. is
supported by UKRI Future Leader Fellowship MR/T019956/1. A.P. additionally
received funding from the European Research Council (ERC) under the
European Union’s Horizon 2020 research and innovation programme under grant
agreements No 757646 \& 813942. N.~H.--T. is funded by the Albert Einstein
Center for Fundamental Physics at the University of Bern.

\bibliographystyle{JHEP}
\bibliography{klfv}

\end{document}